\documentstyle[pra,aps,eqsecnum,psfig]{revtex}
\tighten

\newcommand{\ket}[1]{\left | #1 \right \rangle}
\newcommand{\bra}[1]{\left \langle #1 \right |}

\newcommand{\proj}[1]{\ket{#1} \bra{#1}}
\newcommand{\tr}{{\rm \, Tr }\, }
\begin{document}
\draft

\title{Optimal phase estimation and square root measurement}
\author{Masahide Sasaki$^1$, Alberto Carlini$^{1}$, and
Anthony Chefles$^2$}
\address{${}^1$
   Communications Research Laboratory,
   Ministry of Posts and Telecommunications\\
   Koganei, Tokyo 184-8795, Japan\\
   E-mail:psasaki@crl.go.jp}
\address{${}^2$
   Department of Physical Sciences,
   University of Hertfordshire,\\
   Hatfield AL10 9AB, Herts, UK}


\maketitle

\begin{abstract}
We present an optimal strategy having finite outcomes for estimating
a single parameter of the displacement operator on an arbitrary finite
dimensional system using a finite number of identical samples.
Assuming the uniform {\it a priori} distribution for the displacement
parameter, an optimal strategy can be constructed by making the
{\it square root measurement} based on uniformly distributed sample
points. This type of measurement automatically ensures the global
maximality of the figure of merit, that is, the so called average score
or fidelity.
Quantum circuit implementations for the optimal strategies are
provided in the case of a two dimensional system.

\end{abstract}



\section{Introduction}\label{introduction}

Quantum state estimation deals with how to estimate unknown
parameters  of  a quantum state as precisely as possible.
This problem was studied extensively in the 1970's  in the context
of the formalism of the probability operator measure (POM)  and of
quantum Bayesian inference.
Basic formulations were already established and optimal strategies
were found in various cases. The relevant works are reviewed in Refs.
\cite{Helstrom_QDET,Holevo_book}.

In the 1990's, the same problem was revisited in a new context
where one is allowed to use finite samples of a quantum system
to be estimated,
while most of earlier works were concerned with the estimation
using a single sample.
Massar and Popescu \cite{Massar95} obtained the optimal strategy
for estimating a quantum pure state of a spin-1/2 system, say,  $\ket\rho$,
from $N$ identically prepared samples
$\ket\rho^{\otimes N}$.
Their strategy is based on the use of an infinite continuous set
of projectors in the Hilbert space of $\ket\rho^{\otimes N}$.
For a given unknown quantum system,
the optimal estimation strategy is not unique. As shown in Ref.
\cite{Holevo_book},
one can always find an optimal strategy consisting of an infinite
continuous set of POMs. On the other hand, for a finite
dimensional system there must exist a discrete and finite POM
achieving the same optimal bound as shown by Derka et al.
\cite{Derka98}.
From the point of view of the physical realization the latter is
preferable, while the former might be a more useful mathematical
tool to derive the maximum attainable average fidelity (which is
a commonly used figure of merit for state estimation).
In Ref. \cite{Derka98}
an algorithm is described for constructing such optimal and finite
POMs for an arbitrary finite dimensional system in a pure state.
Latorre et al. \cite{Latorre98} then studied the optimal strategy
with a minimum number of outcomes for a spin-1/2 system,
and showed explicit forms of optimal minimal measurements for
$N=1\sim5$.
Their analysis was extended to the cases of a mixed state of a
spin-1/2 system
\cite{Vidal99},
and of an arbitrary spin system in a pure state
\cite{Acin99}.
In Ref. \cite{Vidal99},
the closed form expressions for the maximum average score,
the optimal minimal POM and its number of outcomes were derived
by using the symmetric properties of the totally symmetric
subspace supported by a tensor product of $N$ identical samples.
To construct the optimal minimal strategy explicitly, however,
some parameters are to be determined and remain unsolved for
larger $N(\ge6)$.
For higher dimensional systems it becomes more difficult to find
concrete forms for the optimal minimal strategy
\cite{Acin99}.
As for the maximum average fidelity the explicit expression for
an arbitrary $N$ was obtained
in Ref. \cite{Bruss99}.
This bound was derived by using the fact that quantum optimal
state estimation using $N$ samples can be viewed as the limiting
case $M\rightarrow\infty$ of universal optimal cloning generating
$M$ copies from $N$ inputs for which the maximum average
fidelity was given by Werner
\cite{Werner98}.
\footnote{
The optimal cloning map of Ref. \cite{Werner98} may have
a connection with the infinite continuous version of
the optimal state estimation strategy.}

Although a substantial progress has been obtained in the
quantum state estimation,
it is still a difficult and open problem how to find explicit
and physically realizable solutions for optimal strategies in an
algorithmic way, especially in the case of higher dimensional
systems and larger numbers of samples.
Moreover, discussions given so far in the literatures
for ensuring the optimality of discrete and finite POMs were
focused only on the condition for extremality and not on the
full conditions for the existence of a global maximum,
which are reviewed in Refs.
\cite{Helstrom_QDET,Holevo_book}.
In general,
seeking all extrema and picking up the point corresponding to
the global maximum is not necessarily a trivial task for
complex systems.

In this paper, we focus on a single parameter estimation of an
arbitrary finite dimensional system in a pure state and give finite
element optimal strategies that can be constructed in a
straightforward way and ensure the global maximality conditions
for the POM.

\section{Optimal phase estimation}\label{sec2}

Consider a finite dimensional system described in a Hilbert
space $\cal H$ and let $\{\ket{0}, \ket{1}, ..., \ket{K} \}$ be
its basis built from the eigenstates of the observable $\hat O$
on $\cal H$, $\hat O\ket{k}=k\ket{k}$.
Such a system may be, e.g.,  an optical field produced by
quantum scissors
\cite{Barnett2000}
(with $\hat O$ the photon number operator)
or
a spin $j$ system described by a superposition of the eigenstates
$\{\ket{j, m} , m=-j, ..., j\}$ (with $\hat O$ the spin operator).
The problem we consider is the estimation of a unitary
evolution $\hat u$ specified by a displacement parameter
$\theta$, that is,
$\hat u(\theta)={\rm e}^{-i\theta\hat O}$.
We suppose that the initial state of the system is {\it a priori}
known and reads
\begin{equation}
\ket{\psi(0)}=\sum_{k=0}^{K} c_k\ket{k}
\label{initial state of a system}
\end{equation}
(with $c_k$ non zero arbitrary complex coefficients),
but we do not have any {\it a priori} knowledge about $\theta$.
After the evolution, the system will be in a state
\begin{equation}
\ket{\psi(\theta)}=\hat u(\theta)\ket{\psi(0)}
                 =\sum_{k=0}^K c_k{\rm e}^{-i\theta k}\ket{k}.
\label{final state}
\end{equation}
It is assumed that $N$ identical samples of the system are
available. The combined system is then described on the
{\it totally symmetric bosonic subspace} of
${\cal H}^{\otimes N}$
\cite{Derka98,Werner98} as
\begin{equation}
\ket{\Psi(\theta)}=\ket{\psi(\theta)}^{\otimes N}
={\sum_{\vec n}}'C(\vec n)
                {\rm e}^{-i\theta\sum_k k n_k}\ket{\vec n},
\label{initial state}
\end{equation}
where
\begin{equation}
C(\vec n)\equiv\sqrt{N!}\prod_{k=0}^K
                {{c_k^{n_k}}\over{\sqrt{n_k!}}},
\label{coef C(n)}
\end{equation}
$\sum_{\vec n}'$ means the summation over $(K+1)$-tuples
$\vec n\equiv(n_0,...,n_K)$ with $\sum_{k=0}^{K} n_k=N$, and
$\ket{\vec n}$ ($\equiv\ket{n_0,...,n_K}$) is
the occupation number basis.
The dimensionality $D_{\rm B}$ of this space is
$D_{\rm B}=\left(\begin{array}{c} N +K\cr K \end{array}\right)$.
For our present purpose, however, it is enough to consider the
smaller subspace spanned by the eigenstates of the compound
operator
$\hat O_{\rm T}\equiv\sum_{i=1}^N \hat O(i)$, where $\hat O(i)$
is the observable for the $i$-th sample.
Let $\{\vec n_i^{(J)}\}$ be the set of $(K+1)$-tuples that satisfy
$\sum_{k=0}^K k n_k=J$ and define
$A_J\equiv\sqrt{\sum_i\vert C(\vec n_i^{(J)})\vert^2}$.
Then the state in Eq. (\ref{initial state}) can be rewritten as
\begin{equation}
\ket{\Psi(\theta)}=\sum_{J=0}^{KN}A_J
                {\rm e}^{-i\theta J}\ket{J},
\label{initial state T}
\end{equation}
where
\begin{equation}
\ket{J}=A_J^{-1}\sum_i C(\vec n_i^{(J)})\ket{n_i^{(J)}}, \quad
\hat O_{\rm T}\ket{J}=J\ket{J}.
\label{eigenstate J}
\end{equation}
The POM describing the optimal estimation strategy is
constructed in the $D_{\rm T}=KN+1$ dimensional subspace
${\cal H}_{\rm T}$ spanned by the set $\{\ket{J}\}$.

Such a POM $\{\hat\mu_m\}$ should maximize the following
score
\begin{equation}
\bar S(N)=\sum_m{1\over{2\pi}}\int_0^{2\pi} d\theta
                 \tr{\left(\hat\mu_m\hat\Psi(\theta)\right)}
                 \vert\langle\psi_m\vert\psi(\theta)\rangle\vert^2,
\label{score}
\end{equation}
where $\hat\Psi(\theta)=\proj{\Psi(\theta)}$ and
$\ket{\psi_m}$ is a guessed state according to the $m$-th
outcome of the measurement.
Optimality can be discussed along with the conditions for
quantum Bayesian optimization
\cite{Helstrom_QDET,Holevo_book}.
Let us introduce the {\it score operators}
\begin{equation}
\hat W_m\equiv{1\over{2\pi}}\int_0^{2\pi} d\theta
                    \hat\Psi(\theta)
                    \vert\langle\psi_m\vert\psi(\theta)\rangle\vert^2.
\label{score_op_def}
\end{equation}
They include all the {\it a priori} information.
Then the necessary and sufficient conditions such that a certain
set $\{\hat\mu_m\}$ globally maximizes the average score
for a fixed set of $\{\ket{\psi_m}\}$ are expressed as
\cite{Helstrom_QDET,Holevo_book}
\begin{equation}
\begin{array}{cl}
\mbox{(i)}&\hat\Gamma\equiv
     \sum_m \hat W_m\hat\mu_m
\mbox{ is hermitian and }
(\hat\Gamma-\hat W_m)\hat\mu_m=0, \quad\forall m, \\
{}&{}\\
\mbox{(ii)}&\hat\Gamma-\hat W_m\ge0, \quad\forall m,
\end{array}
\label{opt_cond}
\end{equation}
where $\hat\Gamma$ is called the Lagrange operator,
and the average score (2.7) can then be rewritten as
\begin{equation}
{\bar S }(N)=\tr{\hat \Gamma}.
\label{av-score-2}
\end{equation}
The optimal estimation strategy can be constructed in the
following way.
First take $M$ states corresponding to uniformly distributed
sample points in $\theta\in[0, 2\pi)$,  that is,
\begin{equation}
\ket{\psi_m}=\sum_{k=0}^K c_k
                     {\rm e}^{-i{{2\pi m}\over M}k}\ket{k},
\quad (m=0, .., M-1),
\label{templates}
\end{equation}
and let us denote its $N$ tensor product states as
$\ket{\Psi_m}=\ket{\psi_m}^{\otimes N}$.
Then consider the state vector
\begin{equation}
\ket{\mu_m}\equiv\hat\Psi^{-{1\over2}}\ket{\Psi_m}, \quad
\hat\Psi\equiv\sum_m\proj{\Psi_m}.
\label{srm_vector}
\end{equation}
The set $\{\hat\mu_m\equiv\proj{\mu_m}\}$ is easily seen to
be a set of non-negative Hermitian operators
satisfying the resolution of the identity on ${\cal H}_T$, and
thus a POM. This POM is often called the
{\it square root measurement}
\cite{Holevo_SubOptMeas78,Hausladen_SubOptMeas95,BanKurokawa_SqRt97}.
If we take $M\ge KN+1$ sample points, $\{\hat\mu_m\}$ also
works as the optimal estimation strategy.
Under the condition $M\ge KN+1$ we have, in fact,
that (cf. Eq. (\ref{initial state T}) with $\theta=2\pi m/M$)
\begin{equation}
\hat\Psi=M\sum_{J=0}^{KN}A_J^2\proj{J},
\end{equation}
because
\begin{equation}
\sum_{m=0}^{M-1}e^{i{{2\pi m}\over M}n}=M\delta_{n,0},
~~\mbox{for}-KN\le n\le KN.
\label{orthogonality}
\end{equation}
Therefore, from Eq. (\ref{srm_vector}) we get
\begin{equation}
\ket{\mu_m}=
        {1\over{\sqrt M}}\sum_{J=0}^{KN}
        {\rm e}^{-i{{2\pi m}\over M}J}\ket{J}.
\label{srm}
\end{equation}
To prove optimality, let us first rewrite the score function
as (see Eqs. (\ref{final state}) and (\ref{templates} ))
\begin{equation}
\vert\langle\psi_m\vert\psi(\theta)\rangle\vert^2
=d_0
+\sum_{L=1}^K d_L
   \left(  {\rm e}^{ i({{2\pi m}\over M}-\theta)L}
           +{\rm e}^{-i({{2\pi m}\over M}-\theta)L}\right),
\label{score_func}
\end{equation}
where
\begin{equation}
d_L=\sum_{k=0}^{K-L}\vert c_{k+L}c_k\vert^2.
\end{equation}
By substituting Eq. (\ref{score_func}) into
Eq. (\ref{score_op_def}), we obtain
\begin{equation}
\hat W_m=d_0\sum_{J=0}^{KN}
                       A_J^2 \vert J\rangle\langle J\vert
               +\sum_{L=1}^K d_L
                 \sum_{J=0}^{KN-L}
                            A_J A_{J+L}\left (
			    {\rm e}^{i{{2\pi m}\over M}L}
                                      \vert J\rangle\langle J+L\vert
                                  +{\rm e}^{-i{{2\pi m}\over M}L}
                                      \vert J+L\rangle\langle J\vert
                           \right ).
\label{score_op_matrix}
\end{equation}
We then have
\begin{equation}
\hat\Gamma=\sum_{m=0}^{M-1} \hat W_m\hat\mu_m
 =d_0\sum_{J=0}^{KN}
                 A_J^2 \vert J\rangle\langle J\vert
    + \sum_{L=1}^K d_L
                 \sum_{J=0}^{KN-L}
                 A_J A_{J+L}
                          \Big(  \vert J\rangle\langle J\vert
                                  +\vert J+L\rangle\langle J+L\vert
                           \Big),
\label{Lagrange_op_matrix}
\end{equation}
where the orthogonality relation Eq. (\ref{orthogonality}) was
used under the condition $M\ge KN+1$.
The operator $\hat\Gamma-\hat W_m$ is now
\begin{equation}
\hat\Gamma-\hat W_m=\sum_{L=1}^K d_L\sum_{J=0}^{KN-L}
       A_J A_{J+L}
                     \biggl [  \vert J\rangle\langle J\vert
                           +\vert J+L\rangle\langle J+L\vert
           -\left ({\rm e}^{ i{{2\pi m}\over M}L}
                                      \vert J\rangle\langle J+L\vert
              +{\rm e}^{-i{{2\pi m}\over M}L}
                                      \vert J+L\rangle\langle J\vert
                                      \right )
\biggr ].
\label{Gamma-Wm}
\end{equation}
Now look at each operator enclosed by $[...]$ in Eq. (\ref{Gamma-Wm}).
This $2\times2$
matrix has the eigenvalues 0 and $2$ and
is non-negative.
So is the sum of them.
Since the coefficients $d_L$ and $A_J$ in
Eq. (\ref{Gamma-Wm}) are also positive,
the second condition (ii) of Eq. (\ref{opt_cond}) is proved.
The first condition (i) can be easily checked by direct calculation.
The maximum average score is then given by
\begin{equation}
{\bar S}_{\rm MAX}(N)
     =d_0\sum_{J=0}^{KN} A_J^2
     + 2\sum_{L=1}^K d_L \sum_{J=0}^{KN-L} A_J A_{J+L}.
\label{maximum_score}
\end{equation}
This maximum is independent of the number of sample points $M$,
that is, for $M\ge KN+1$ the attainable average score is saturated.
In fact, in the limit of $M\rightarrow\infty$ we can construct
the infinite continuous POM
\begin{equation}
d\hat\Pi(\varphi)
\equiv\proj{\mu(\varphi)}{d\varphi}/{2\pi}, ~~
\int_0^{2\pi}d\hat\Pi(\varphi)=\hat I,
\label{contPOM}
\end{equation}
where
\begin{eqnarray}
\ket{\mu(\varphi)}&\equiv&
         \left(  {1\over{2\pi}}\int_0^{2\pi} d\phi
                   \hat\Psi(\phi)
\right)^{-{1\over2}}\ket{\Psi(\varphi)}
\nonumber \\
       &=&\sum_{J=0}^{KN}
        {\rm e}^{-i\varphi J}\ket{J},
\label{srm_continuous}
\end{eqnarray}
and this attains the same maximum as Eq. (\ref{maximum_score}).
Thus it is proved that the measurement state vector
Eq. (\ref{srm_vector}) (Eq. (\ref{srm})) provides the optimal
estimation
strategy. For the minimum number of sample points $M=KN+1$,
$\{\ket{\mu_m}\}$ is an orthonormal set, that is,
a von Neumann measurement.

Here we mention other strategies $\{{\hat \mu}^{\bot}_m\}$
which extremize the average score Eq. (\ref{score})
(for the same ${\hat W}_m$ as in Eq. (\ref{score_op_def})),
that is,
satisfy the first condition  (i) of Eq. (\ref{opt_cond}),
but not necessarily the second condition (ii).
Consider, for example, the states orthogonal to
the $N$-tensor-product sample states
$\ket{\Psi_m}$,
that is, the states
\begin{equation}
\ket{\Psi_m^\bot}=\sum_{J=0}^{KN}
                \left(\begin{array}{c} KN \cr J \end{array}\right)
                A_J^{-1}(-1)^J
                {\rm e}^{-i  {{2\pi m}\over M}  J}\ket{J}.
\label{reciprocal state}
\end{equation}
The square root measurement for discriminating
$\{\ket{\Psi_m^\bot}\}$ is given as
\begin{eqnarray}
\ket{\mu_m^\bot}
       &\equiv&\Big(
                              \sum_{m=0}^{M-1} \proj{\Psi_m^\bot}
                      \Big)^{-{1\over2}}\ket{\Psi_m^\bot} \nonumber\\
       &=&{1\over{\sqrt M}}
             \sum_{J=0}^{KN}
             (-1)^J {\rm e}^{-i{{2\pi m}\over M}J}\ket{J}.
\label{recipro_srm_vector}
\end{eqnarray}
The Lagrange operator for this measurement is
\begin{equation}
\hat\Gamma^\bot
  =\sum_{m=0}^{M-1} \hat W_m\hat\mu_m^\bot
  =d_0\sum_{J=0}^{KN}
                 A_J^2 \vert J\rangle\langle J\vert
    + \sum_{L=1}^K d_L
                 \sum_{J=0}^{KN-L}
                 A_J A_{J+L}(-1)^L
                          \Big(  \vert J\rangle\langle J\vert
                                  +\vert J+L\rangle\langle J+L\vert
                           \Big),
\label{Lagrange_op_recipro}
\end{equation}
and we have that
\begin{eqnarray}
\hat\Gamma^\bot-\hat W_m&=&\sum_{L=1}^K d_L\sum_{J=0}^{KN-L}
       A_J A_{J+L}
       \biggl [(-1)^L(  \vert J\rangle\langle J\vert
                           +\vert J+L\rangle\langle J+L\vert  \big)
\nonumber\\
           &-& \left ({\rm e}^{ i{{2\pi m}\over M}L}
                                      \vert J\rangle\langle J+L\vert
              +{\rm e}^{-i{{2\pi m}\over M}L}
                                      \vert J+L\rangle\langle J\vert
                                      \right )
\biggr ],
\label{Gamma-Wm_recipro}
\end{eqnarray}
which is easily seen to satisfy
\begin{equation}
(\hat\Gamma^\bot-\hat W_m)\ket{\mu_m^\bot}=0.
\end{equation}
Thus the POM $\{\hat\mu_m^\bot\equiv\proj{\mu_m^\bot}\}$ is an
extremal solution for the average score Eq. (\ref{score}).
However, e.g., in the case of a two dimensional system
($K=1$ in Eq. (\ref{initial state of a system})),
Eq. (\ref{Gamma-Wm_recipro}) reduces to
\begin{equation}
\hat\Gamma^\bot-\hat W_m=-d_1\sum_{J=0}^{N-1}
       A_J A_{J+1}
                     \biggl [ \vert J\rangle\langle J\vert
                           +\vert J+1\rangle\langle J+1\vert
+{\rm e}^{ i{{2\pi m}\over M}}
                                      \vert J\rangle\langle J+1\vert
             +{\rm e}^{-i{{2\pi m}\over M}}
                                      \vert J+1\rangle\langle J\vert
\biggr ],
\label{Gamma-Wm_recipro_2dim}
\end{equation}
which is clearly a non-positive definite operator, that is,
$\hat\Gamma^\bot-\hat W_m\le0$.
In contrast to Eqs. (\ref{opt_cond}),
this means that
$\{\hat\mu_m^\bot\}$ attains a {\it global minimum} of
the average score.
In the general case of $K>1$, it is not necessarily the case that
$\hat\Gamma^\bot-\hat W_m$ is positive or negative definite.
Thus, in general, the set $\{\hat\mu_m^\bot\}$ may represent
strategies that attain either {\it local} maxima or minima.

\section{Quantum circuit for the optimal strategy}\label{circuit}

Let us consider physical implementations of
the optimal estimation strategy represented by Eq. (\ref{srm}).
This is a collective measurement on a finite sample system.
This sort of measurement can, in principle, be realized by
a quantum circuit acting on the combined system and
a separable measurement on each sample system.
From this point of view, the main problem is a synthesis of
an appropriate quantum circuit.
When we take the minimum number of outputs $M=KN+1$,
the measurement basis $\{\ket{\mu_m}\}$ is orthonormal,
and Eq. (\ref{srm}) is the discrete Fourier transform in
the subspace ${\cal H}_{\rm T}$.
The discrete Fourier transform is a fundamental tool in
quantum computation.
The corresponding circuit working on qubit systems is
already known (see, e.g., Ref.
\cite{Cleve_QAlgorith98}).
Therefore in the case of a two dimensional system
(that is, $K=1$ in Eq. (\ref{final state})),
we may apply this result to synthesizing the optimal
estimation strategy.

\begin{figure}[htb]
\centerline{\psfig{file=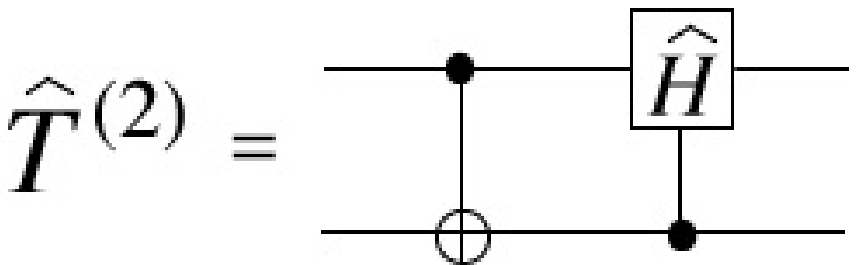,width=5cm}}
\caption{The circuit for the basis transformation from
$\{\ket{0}_{\rm T}, \ket{1}_{\rm T}, \ket{2}_{\rm T}\}$
to $\{ \ket{00}, \ket{01}, \ket{10} \}$.
$\hat H$ is the Hadamard transformation. }
\label{T_2}
\end{figure}

\begin{figure}[htb]
\centerline{\psfig{file=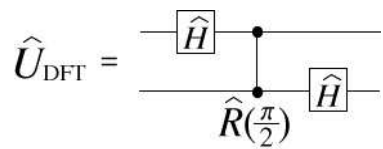,width=5cm}}
\caption{The circuit for the discrete Fourier transform
on two qubit systems. The two bit gate $\hat R(\phi)$
performs the transformation
$\ket{x}\ket{y}\mapsto
{\rm e}^{ixy\phi}\ket{x}\ket{y}$. }
\label{U_DFT}
\end{figure}

\begin{figure}[htb]
\centerline{\psfig{file=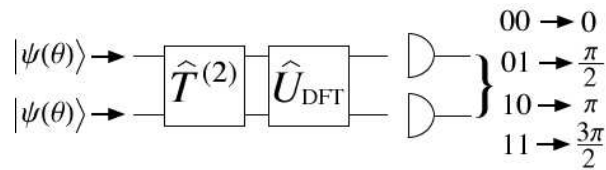,width=8cm}}
\caption{The circuit structure for the optimal estimation
strategy in the case of $N=2$. }
\label{circuit_N=2}
\end{figure}

Let us start with the simplest two dimensional system
with $N=2$.
The state to be measured is
\begin{equation}
\ket{\psi(\theta)}^{\otimes2}
     = A_0 \ket{0}_{\rm T}
     + A_1 {\rm e}^{-i\theta} \ket{1}_{\rm T}
     + A_2 {\rm e}^{-i2\theta} \ket{2}_{\rm T},
\end{equation}
where $A_0=c_0^2$, $A_1={\sqrt2}c_0 c_1$, $A_2=c_1^2$,
and
\begin{mathletters}
\begin{eqnarray}
\ket{0}_{\rm T}&=&\ket{00}, \\
\ket{1}_{\rm T}&=&{1\over{\sqrt2}}
                              \left(\ket{01}+\ket{10}\right),\\
\ket{2}_{\rm T}&=&\ket{11},
\end{eqnarray}
\end{mathletters}

\noindent
assuming that the coefficients $c_i$ are real for
eliminating unessential phase factors.
Here we have used the subscript T for denoting the basis
$\ket J$ of the $KN+1$ dimensional subspace
${\cal H}_{\rm T}$, and
the basis states $\ket{00}$, $\ket{01}$, $\ket{10}$ and $\ket{11}$
span the space ${\cal H}^{\otimes2}$.
The optimal strategy with the minimal outputs is represented
as
\begin{equation}
\ket{\mu_m}={1\over{\sqrt3}}
                    \left(              \ket{0}_{\rm T}
                    + {\rm e}^{-i{{2\pi m}\over3}} \ket{1}_{\rm T}
                    + {\rm e}^{-i{{4\pi m}\over3}} \ket{2}_{\rm T}
                    \right), \quad (m=0, 1, 2).
\end{equation}
This expression is, however, rather inconvenient for directly
applying the discrete Fourier transform quantum network, which is
usually defined in a $2^n$ dimensional space corresponding
to $n$ qubit systems.
Therefore, it seems better to consider the measurement
circuit in a 4 dimensional space ${\cal H}^{\otimes2}$ spanned by
the basis $\{ \ket{00}, \ket{01}, \ket{10}, \ket{11} \}$.
In fact we also have the 4-output optimal, nonminimal ($M=4$) strategy in
${\cal H}_{\rm T}$ as
\begin{equation}
\ket{\mu_m}={1\over{\sqrt4}}
                    \left(              \ket{0}_{\rm T}
                    + {\rm e}^{-i{{2\pi m}\over4}} \ket{1}_{\rm T}
                    + {\rm e}^{-i{{4\pi m}\over4}} \ket{2}_{\rm T}
                    \right), \quad (m=0, 1, 2, 3),
\end{equation}
and
this can be extended to a von Neumann
measurement in ${\cal H}^{\otimes2}$.
It is thus helpful to transform the state of the
input samples by the unitary operator $\hat T^{(2)}$ defined
by the circuit shown in Fig. \ref{T_2} such that
\begin{equation}
\hat T^{(2)}\ket{\psi(\theta)}^{\otimes2}
     = A_0 \ket{00}
     + A_1 {\rm e}^{-i\theta} \ket{01}
     + A_2 {\rm e}^{-i2\theta} \ket{10}.
\end{equation}
We may then apply the measurement
\begin{equation}
\ket{\mu_m}={1\over{\sqrt4}}
                    \left(              \ket{00}
                    + {\rm e}^{-i{{2\pi m}\over4}} \ket{01}
                    + {\rm e}^{-i{{4\pi m}\over4}} \ket{10}
                    \right), \quad (m=0, 1, 2, 3),
\end{equation}
or its Naimark extension in ${\cal H}^{\otimes2}$
\begin{equation}
\ket{\Pi_m}={1\over{\sqrt4}}
                    \left(              \ket{00}
                    + {\rm e}^{-i{{2\pi m}\over4}} \ket{01}
                    + {\rm e}^{-i{{4\pi m}\over4}} \ket{10}
                    + {\rm e}^{-i{{6\pi m}\over4}} \ket{11}
                    \right), \quad (m=0, 1, 2, 3)
\label{Pi_m}
\end{equation}
which, using the discrete Fourier transform
$\hat U_{\rm DFT}$ shown by the circuit of
Fig. \ref{U_DFT}, can be explicitly written as
\begin{mathletters}
\begin{eqnarray}
\ket{\Pi_0}&=&\hat U_{\rm DFT}^\dagger\ket{00},\\
\ket{\Pi_1}&=&\hat U_{\rm DFT}^\dagger\ket{10},\\
\ket{\Pi_2}&=&\hat U_{\rm DFT}^\dagger\ket{01},\\
\ket{\Pi_3}&=&\hat U_{\rm DFT}^\dagger\ket{11}.
\end{eqnarray}
\end{mathletters}
\noindent
Thus the optimal phase estimation can be realized
by first performing the unitary transformation
$\hat U_{\rm DFT}\hat T^{(2)}$ on the input state
$\ket{\psi(\theta)}^{\otimes2}$, then measuring
the transformed state in the basis
$\{ \ket{00}, \ket{10}, \ket{01}, \ket{11} \}$
(which is a separable measurement), and
finally deciding the phase as
$\theta=0, {\pi\over2}, \pi$, or ${{3\pi}\over2}$,
according to whether the outcome is $\ket{00}$, $\ket{10}$, $\ket{01}$, or
$\ket{11}$,
respectively.  This is summarized in Fig.
\ref{circuit_N=2}.

Similarly, in the case of $N=3$, the input state is represented as
\begin{equation}
\ket{\psi(\theta)}^{\otimes3}
     = A_0 \ket{\bar 0}_{\rm T}
         + A_1 {\rm e}^{-i\theta} \ket{\bar 1}_{\rm T}
         + A_2 {\rm e}^{-i2\theta} \ket{\bar 2}_{\rm T}
         + A_3 {\rm e}^{-i3\theta} \ket{\bar 3}_{\rm T},
\end{equation}
where
\begin{mathletters}
\begin{eqnarray}
\ket{\bar 0}_{\rm T}&=&\ket{000}, \\
\ket{\bar 1}_{\rm T}&=&{1\over{\sqrt3}}
           \big(\ket{001}+\ket{010}+\ket{100}\big),\\
\ket{\bar 2}_{\rm T}&=&{1\over{\sqrt3}}
           \big(\ket{110}+\ket{101}+\ket{011}\big),\\
\ket{\bar 3}_{\rm T}&=&\ket{111}.
\end{eqnarray}
\end{mathletters}

\begin{figure}[htb]
\centerline{\psfig{file=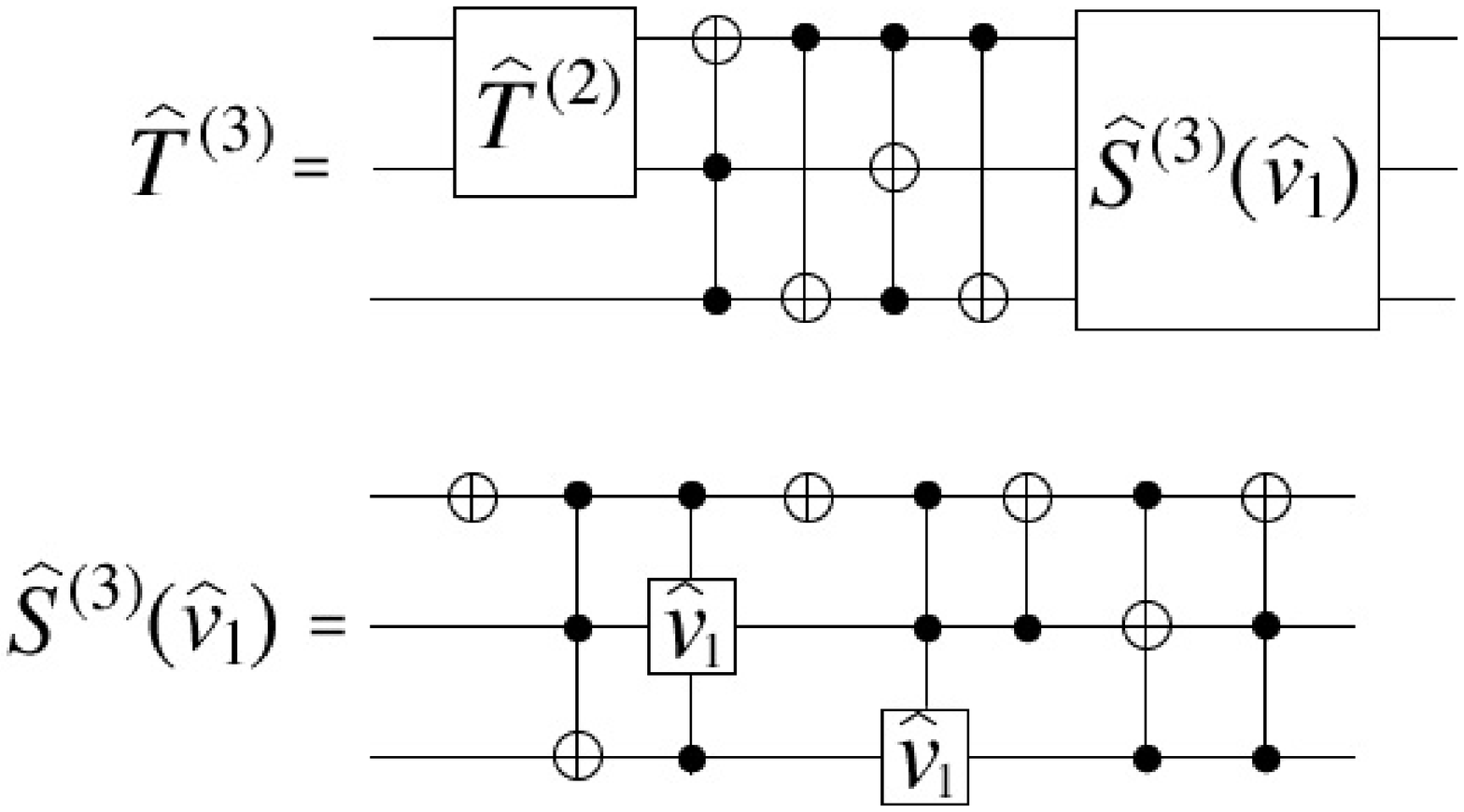,width=7.3cm}}
\caption{The circuit which converts the basis states
$\{ \ket{\bar 0}_{\rm T}, \ket{\bar 1}_{\rm T},
      \ket{\bar 2}_{\rm T}, \ket{\bar 3}_{\rm T} \}$
into $\ket{0}\otimes  \{ \ket{00}, \ket{01}, \ket{10}, \ket{11} \}$. }
\label{T_3}
\end{figure}

\noindent
Let $\hat T^{(3)}$ be the unitary operator
which converts the basis states
$\{ \ket{\bar 0}_{\rm T}, \ket{\bar 1}_{\rm T},
      \ket{\bar 2}_{\rm T}, \ket{\bar 3}_{\rm T} \}$
into $\ket{0}\otimes \{ \ket{00}, \ket{01}, \ket{10}, \ket{11} \}$,
respectively, that is
\begin{equation}
\hat T^{(3)}\ket{\psi(\theta)}^{\otimes3}
      = \ket{0}\otimes\left(
          A_0 \ket{00}
       + A_1 {\rm e}^{-i\theta} \ket{01}
       + A_2 {\rm e}^{-i2\theta} \ket{10}
       + A_3 {\rm e}^{-i3\theta} \ket{11}
                              \right).
\end{equation}
The estimation strategy can then be constructed again
in the 4 dimensional space ${\cal H}^{\otimes2}$,
where
the minimal optimal measurement is actually given by
Eq. (\ref{Pi_m})
and can be realized just as in the previous case.
The unitary operator $\hat T^{(3)}$ can be effected by the
circuit shown in Fig. \ref{T_3} which consists of three
main blocks.
In the first block, the operator $\hat T^{(2)}$ acts on the
first two qubits of the ${\cal H}_T$ basis states, which gives
\begin{mathletters}
\begin{eqnarray}
\ket{\bar 0}_{\rm T}&\mapsto&\ket{000},
\label{basis_a}\\
\ket{\bar 1}_{\rm T}&\mapsto&
                     {1\over\sqrt3}\ket{001}
                   +\sqrt{2\over3}\ket{010},
\label{basis_b}\\
\ket{\bar 2}_{\rm T}&\mapsto&
                     {1\over\sqrt3}\ket{100}
                   +\sqrt{2\over3}\ket{011},
\label{basis_c}\\
\ket{\bar 3}_{\rm T}&\mapsto&\ket{101}.
\label{basis_d}
\end{eqnarray}
\end{mathletters}
The second block transforms the state $\ket{100}$ into the state
$\ket{110}$ and the last block $\hat S^{(3)}(\hat v_1)$,
which includes the conditional rotation on one qubit
\begin{equation}
\hat v_1=\left(
               \begin{array}{cc}
               {1\over\sqrt3}&\sqrt{2\over3}\\
             -\sqrt{2\over3}&{1\over\sqrt3}
               \end{array}
               \right),
\end{equation}
leads to the final basis states $\ket{0}\otimes \{ \ket{00}, \ket{01},
\ket{10}, \ket{11} \}$ as required.

Finally we mention the case of $N=4$.
Let us observe that the ${\cal H}_T$ basis can be
written as
\begin{mathletters}
\begin{eqnarray}
\ket{\bar{\bar 0}}_{\rm T}&=&\ket{000}\otimes\ket{0}, \\
\ket{\bar{\bar 1}}_{\rm T}&=&{1\over2}
      \Big[
      \big(\ket{001}+\ket{010}+\ket{100}\big)
            \otimes\ket{0}
      +\ket{000}\otimes\ket{1}
     \Big]\\
\ket{\bar{\bar 2}}_{\rm T}&=&{1\over\sqrt6}
      \Big[
      \big(\ket{001}+\ket{010}+\ket{100}\big)
            \otimes\ket{1}
      +
      \big(\ket{110}+\ket{101}+\ket{011}\big)
            \otimes\ket{0}
     \Big]\\
\ket{\bar{\bar 3}}_{\rm T}&=&{1\over2}
      \Big[
      \big(\ket{110}+\ket{101}+\ket{011}\big)
            \otimes\ket{1}
      +\ket{111}\otimes\ket{0}
     \Big]\\
\ket{\bar{\bar 4}}_{\rm T}&=&\ket{111}\otimes\ket{1}
\end{eqnarray}
\end{mathletters}

\begin{figure}[htb]
\centerline{\psfig{file=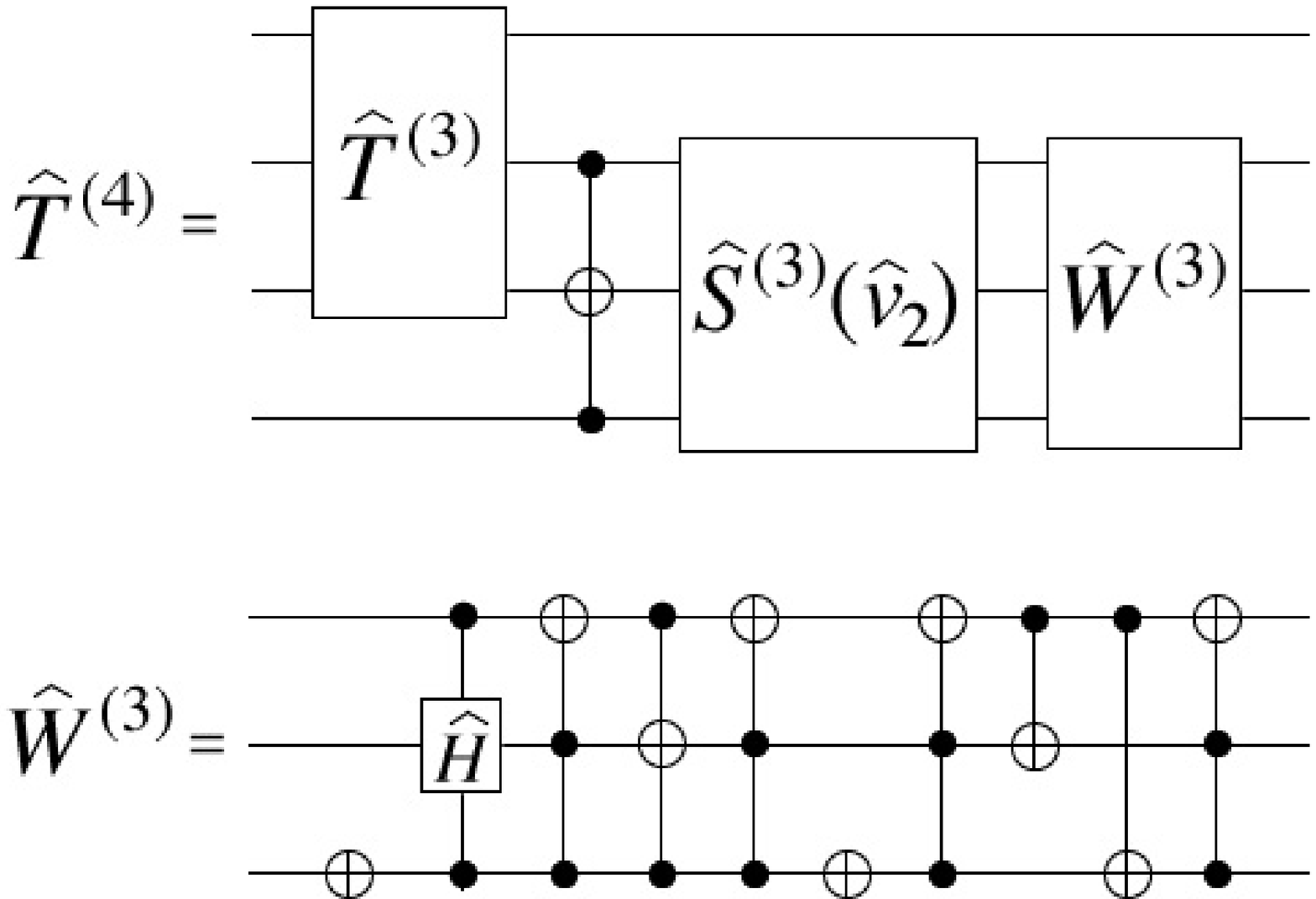,width=7cm}}
\caption{The circuit which transforms the basis  states
$\{ \ket{\bar{\bar 0}}_{\rm T}, ...,
\ket{\bar{\bar 4}}_{\rm T} \}$
into
$\ket{0}\otimes
\{ \ket{000}, \ket{001}, \ket{010},
    \ket{011}, \ket{100} \}$, respectively. }
\label{T_4}
\end{figure}

\noindent
The first 3 qubits can be transformed into
$\ket{000}$, $\ket{001}$, $\ket{010}$, or $\ket{011}$
by applying $\hat T^{(3)}$.
Thus the first qubit of all the basis states becomes $\ket{0}$,
and can be factorized out.
By further applying on the remaining 3 qubits
a CC-NOT gate, the operator
$\hat S^{(3)}(\hat v_2)$ with
\begin{equation}
\hat v_2=\left(
               \begin{array}{cc}
               {1\over2}&{{\sqrt3}\over2}\\
             -{{\sqrt3}\over2}&{1\over2}
               \end{array}
               \right),
\end{equation}
and the operator $\hat W^{(3)}$ as shown in
Fig. \ref{T_4},
the basis states
$\{ \ket{\bar{\bar 0}}_{\rm T}, ...,
\ket{\bar{\bar 4}}_{\rm T} \}$
are finally transformed into
$\ket{0}\otimes
\{ \ket{000}, \ket{001}, \ket{010},
    \ket{011}, \ket{100} \}$, respectively.

\noindent
Let $\hat T^{(4)}$ be this basis transformation represented
by the circuit of Fig. \ref{T_4}.
After transforming the input state by $\hat T^{(4)}$,
it is then sufficient to perform the measurement on
the last 3 qubits.
Let $\{\ket{L}_3; L=0, 1, ..., 7 \}$ be the 3 qubit basis
$\{ \ket{000}, \ket{001}, ..., \ket{111} \}$.
Then the minimal optimal measurement is given as
\begin{equation}
\ket{\mu_m}={1\over{\sqrt5}}
                    \sum_{L=0}^4
                    {\rm e}^{-i{{2\pi m}\over5}L}
                    \ket{L}_3, \quad (m=0, 1, 2, 3, 4).
\end{equation}
However, for applying the discrete Fourier transform
network, it is convenient to take the other optimal
strategy consisting of the overcomplete states
\begin{equation}
\ket{\mu_m}={1\over{\sqrt8}}
                    \sum_{L=0}^4
                    {\rm e}^{-i{{2\pi m}\over8}L}
                    \ket{L}_3, \quad (m=0, 1, ... , 7).
\end{equation}
These can be orthogonalized in the 8 dimensional space
${\cal H}^{\otimes3}$ as
\begin{equation}
\ket{\Pi_m}={1\over{\sqrt8}}
                    \sum_{L=0}^7
                    {\rm e}^{-i{{2\pi m}\over8}L}
                    \ket{L}_3, \quad (m=0, 1, ... , 7).
\end{equation}
This is just the discrete Fourier transform in
${\cal H}^{\otimes3}$, and can be written in the form
$\ket{\Pi_m}=\hat U_{\rm DFT}^\dagger\ket{m}_3$.
The circuit corresponding to $\hat U_{\rm DFT}$ is
found in Ref.
\cite{Cleve_QAlgorith98}.
Therefore the optimal estimation is realized by
applying the unitary transform
$(\hat I\otimes\hat U_{\rm DFT})\cdot\hat T^{(4)}$
on the input state $\ket{\psi(\theta)}^{\otimes4}$
and then by measuring the last 3 qubits of the
transformed state in the basis $\{\ket{L}_3 \}$.
According to the outcome, we decide the phase
to be ${\rm exp}(-i2\pi L/8)$, respectively.

The case of larger $N$ can be treated in a similar
way by applying the circuits used in the case of lower
$N$ inductively.
In the general case of higher dimensional systems,
one should first develope basic tools for costructing
quantum circuits (for some recent progress see, e.g., Ref.
\cite{Alber_generalized_gate2000}), and at present
practical circuit synthesizations remain
an open problem.

\section{Concluding remarks}\label{sec5}

We have shown how to construct the optimal strategies
for estimating a displacement parameter of an arbitrary
finite dimensional system in a pure state.
These are based on the square root measurement for
discriminating the states corresponding to the uniformly
distributed sample points of the parameter.
We have assumed that
{\it a priori} probability distribution of the parameter
is uniform.
When the {\it a priori} distribution is not uniform,
or the system to be estimated is in a mixed state, the
strategy based on the square root measurement is
not in general optimal.

Within the assumption of a uniform {\it a priori}
distribution and the purity of the system to be
estimated, it is a remaining problem whether our
method applies to the estimation of two or more
parameters.
The simplest case would be the estimation of a two
state system using finite identical samples
$\ket{\psi(\theta,\phi)}^{\otimes N}$.
As for the infinite continuous POM,
we can show that the square root measurement
\begin{equation}
d\hat\Pi(\theta,\phi)
\equiv\proj{\mu(\theta,\phi)}
{ {d\phi d\theta{\rm sin}\theta}
  \over
  {2\pi} }, ~~
\int_0^{2\pi}\int_0^{\pi}
d\hat\Pi(\theta,\phi)=\hat I,
\label{contPOM_2param}
\end{equation}
where
\begin{equation}
\ket{\mu(\theta,\phi)}\equiv
  \left[  {1\over{4\pi}}
             \int_0^{2\pi} d\phi'
             \int_0^{\pi} d\theta'{\rm sin} \theta'
             \left( \ket{\psi(\theta,\phi)}
             \bra{\psi(\theta,\phi)} \right) ^{\otimes N}
   \right] ^{-{1\over2}}
   \ket{\psi(\theta,\phi)}^{\otimes N},
\label{srm_cont_2param}
\end{equation}
provides the optimal strategy.
Whether its discrete and finite version should exist
and be built from the uniformly distributed sample
points of $(\theta,\phi)$ is still an open question
at present.

\acknowledgements

The authors would like to thank O. Hirota, S. M. Barnett,
A. S. Holevo, and C. A. Fuchs for valuable discussions.
Alberto Carlini's research is supported by JISTEC under grant n. 199016.
Anthony Chefles's research is supported by EPSRC and the British Council.


\end{document}